\documentclass{aa}

\usepackage{graphicx,natbib}
\bibpunct{(}{)}{;}{a}{}{,}

\def\LB{$\lambda$\,Bootis}
\def\Vmicro{$v_{\rm micro}$} 
\def\Vsini{$v\cdot\sin{i}$}
\def\Teff{$T_{\rm eff}$} 
\def\logg{$\log{g}$}
\def\logZ{[M/H]}

\def\kms{km\,s$^{-1}$}

\begin{document}

   \title{The abundance pattern of the \LB\ stars.
        \thanks{Based on observations obtained at
        the Osservatorio Astronomico di Padua-Asiago, OPD/LNA, KPNO
        and DSO}}

   \author{U. Heiter\inst{1,2}
          }

   \offprints{U. Heiter}

   \institute{Institute for Astronomy (IfA), University of Vienna,
              T\"urkenschanzstrasse 17, A-1180 Vienna\\
              \and
      	      Department of Astronomy, Case Western Reserve University,\\
      	      10900 Euclid Avenue, Cleveland, OH 44106-7215, USA\\
      	      email: ulrike@fafnir.astr.cwru.edu
             }

   \date{Received ; accepted }

\abstract{
Within a project to investigate the properties of \LB\ stars, we
report on their abundance pattern.
High resolution spectra have been obtained for a total of twelve candidate \LB\ stars,
four of them being contained in spectroscopic binary systems,
and detailed abundance analyses have been performed.
All program stars show a characteristic \LB\ abundance pattern 
(deficient heavy elements and solar abundant light elements) 
and an enhanced abundance of Na.
This work raises the fraction of \LB\ stars with known abundances to 50~\%.
The resulting abundances complemented by literature data are used to
construct a ``mean \LB\ abundance pattern'', which exhibits, apart from general
underabundances of heavy elements ($\approx-1$~dex) and solar abundances of
C, N, O, Na and S, a star-to-star scatter which is up to twice as large as 
for a comparable sample of normal stars. 
}

   \maketitle

\section{Introduction}

\LB\ stars are defined as Population\,{\sc i} 
A- to F-type stars, which are metal-poor but exhibit 
nearly solar element abundances for C, N, O and S 
\citep[e.g.][]{Hauc:83,Abt:84b,Gray:97,Paun:00}.
A historical review of studies dealing with these stars, as well
as extensive work on several aspects regarding the \LB\ group,
like spectral classification, photometry, ultraviolet and infrared fluxes,
space motions, binarity and theoretical considerations, can be found
in \citet{Paun:00}. 

From the definition above it is obvious that
the chemical composition of the atmospheres of \LB\ stars represents the most
important property for the characterization
of this group of stars. At the beginning of this project,
 few abundance determinations for \LB\ stars existed in the literature 
(see Sect.~\ref{pattern}). It was not possible to
make a firm conclusion about ``the abundance pattern'' of the \LB\ stars, 
in particular because only a part of the parameter
space occupied by current lists of \LB\ candidates was covered by abundance
analyses.

A detailed knowledge of the common abundance pattern, if existent, 
and of the variation of abundances of individual elements with stellar
parameters is the key to understanding the astrophysical processes 
generating the properties of the \LB\ group.
Therefore the determination of abundances for
as many elements as possible for preferably all members of this group
(where the membership assignment is based on classification spectroscopy)
is regarded as a main goal of this investigation.
Together with all abundance analyses found in the literature, 
the proportion of \LB\ stars with known abundances, based on
the list of \LB\ stars given by \citet{Paun:00},
amounts now to about 50\%, which allows to draw conclusions on the abundance
pattern of the \LB\ stars as a group. A statistical
investigation of the \LB\ parameters and a comparison with other groups
of stars and the interstellar medium will be the topic of a subsequent paper.
 

\begin{table*}
\caption{Observations used in this work. The last three columns give the
observed wavelength range ($\lambda$), the resolving power (R), and the
mean continuum signal to noise ratio determined from the reduced spectra.}
\label{obs_tab}
\begin{tabular}{llclccl}
\hline\hline
HD & Observatory / Telescope & Date & Observer & $\lambda$ [\AA] & R & S/N \\
\hline\hline
74873, 84948, 101108, 106223, & Asiago (Italy) / 1.8 m & 3--1995 & U.\,Heiter, & 4000$-$5700 & 20000 & 150, \\
110411                        &                       &         & E.\,Paunzen &             &       & 250  \\
84948, 171948                 &                       & 2--1997 &             & 4500$-$7200 &       & 200 \\
\hline
107233, 142703, & OPD/LNA (Brazil) / 1.6 m & 6--1995 & E. Paunzen & 3900$-$4900 & 22000 & 150, \\
168740, 170680  &                         &         &            &             &       & 200, 230 \\
\hline
106223 & KPNO (Arizona) / 1.5 m & 4--1998 & M. Weber & 4240$-$4570 & 24000 & 200 \\
\hline
\hline
74873, 101108, 170680 & DSO (N. Carolina) / 0.8 m & 1995 & R.O.\,Gray & 4000$-$4450 & 2200 & 300, 150, 400 \\
\hline
106223 & OHP (France) / 1.93 m & 1995 & E.Paunzen, & 4000$-$4450 & 1200 & 300 \\
\cline{1-3}\cline{5-7}
110411 & Asiago (Italy) / 1.8 m & 1995 & \ U.Heiter & 4000$-$4450 & 1100 & 350 \\
\hline
107233, 142703 & OPD/LNA (Brazil) / 1.6 m & 1995 & E.Paunzen & 4000$-$4450 & 2200 & 200, 350 \\
\hline\hline
\end{tabular}
\end{table*}

\section{Observations and analysis}
\begin{figure*}
\resizebox{\hsize}{!}{\includegraphics[angle=90]{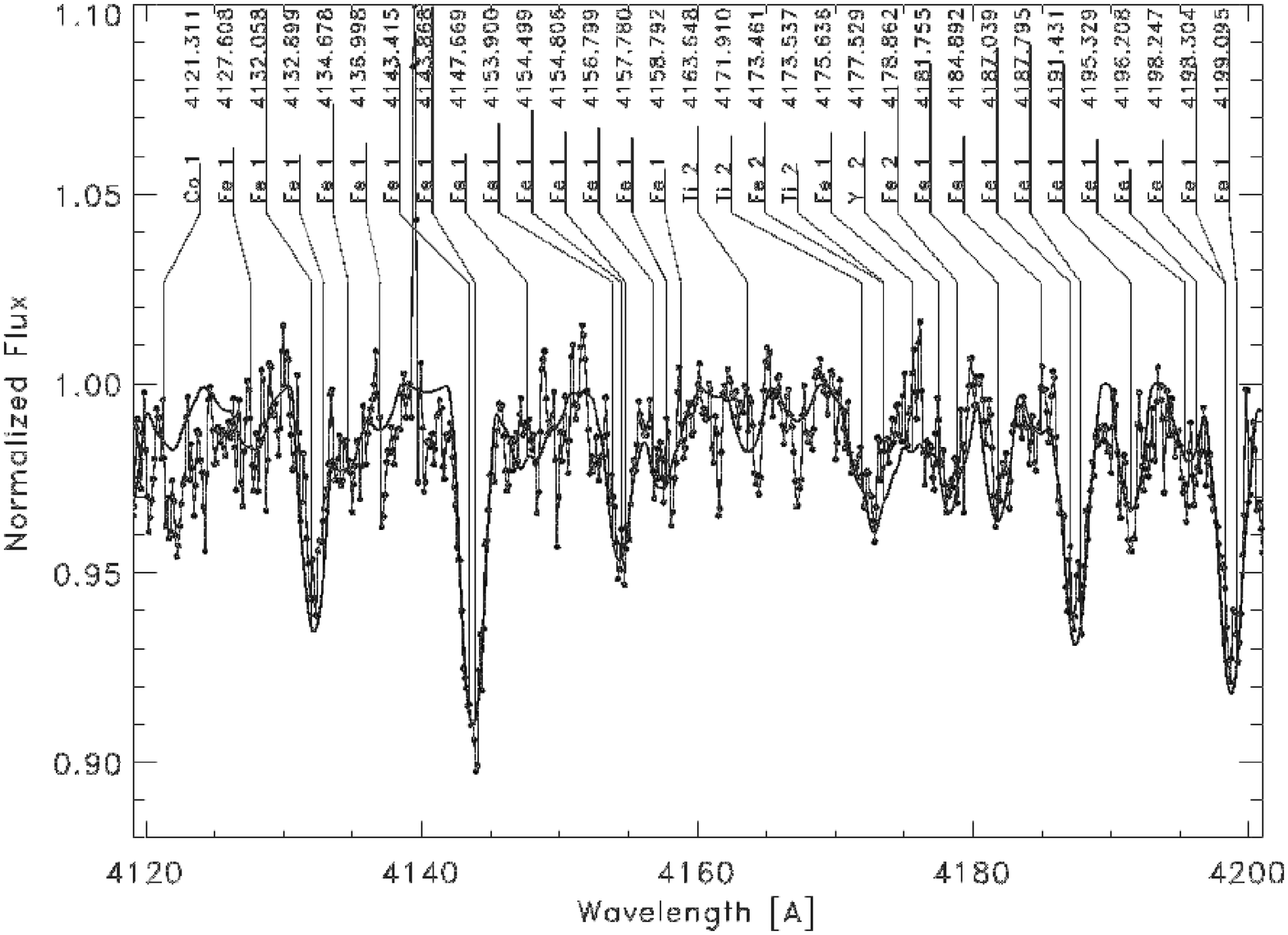}
                      \includegraphics[angle=90]{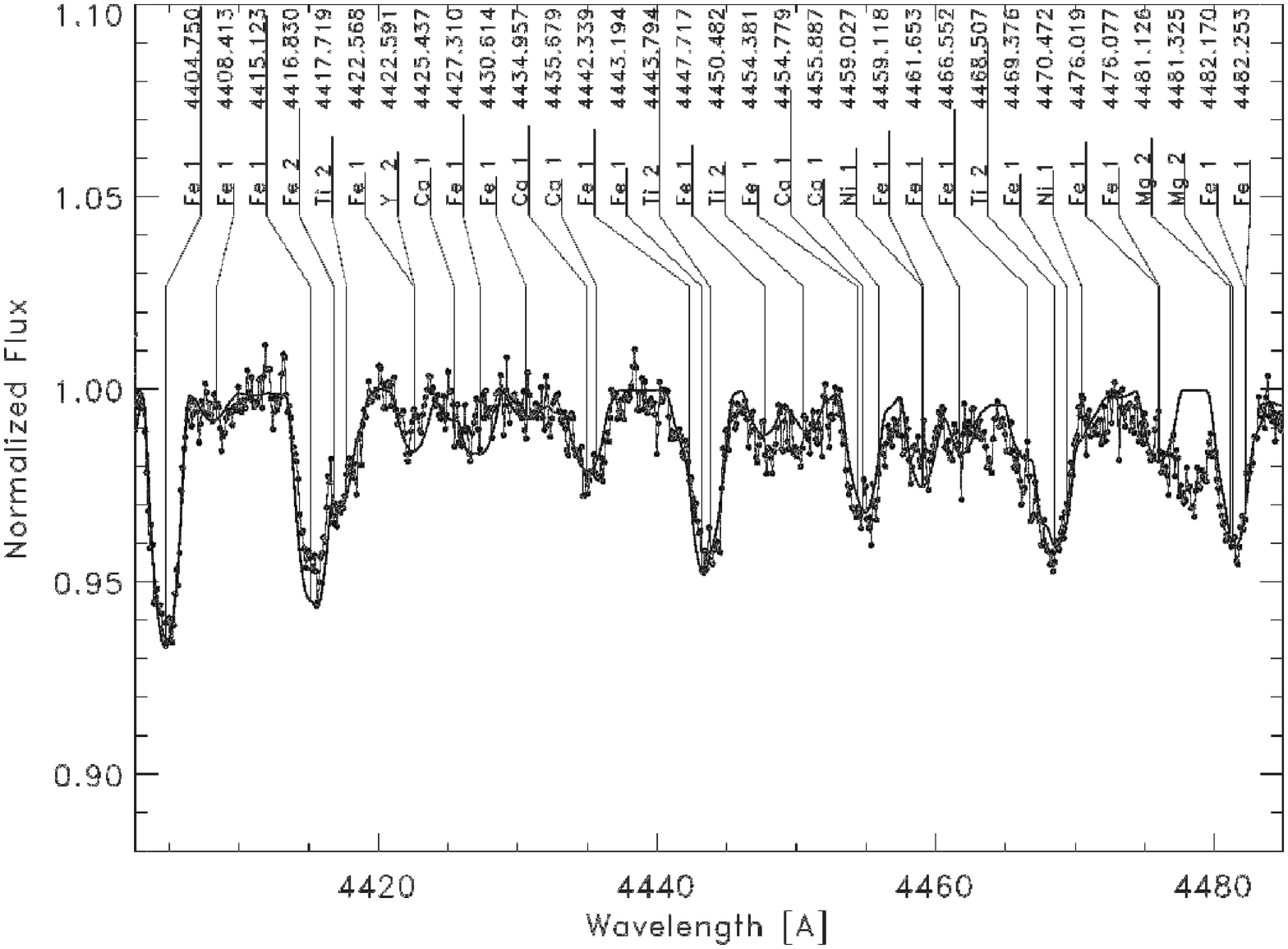}}
\resizebox{\hsize}{!}{\includegraphics[angle=90]{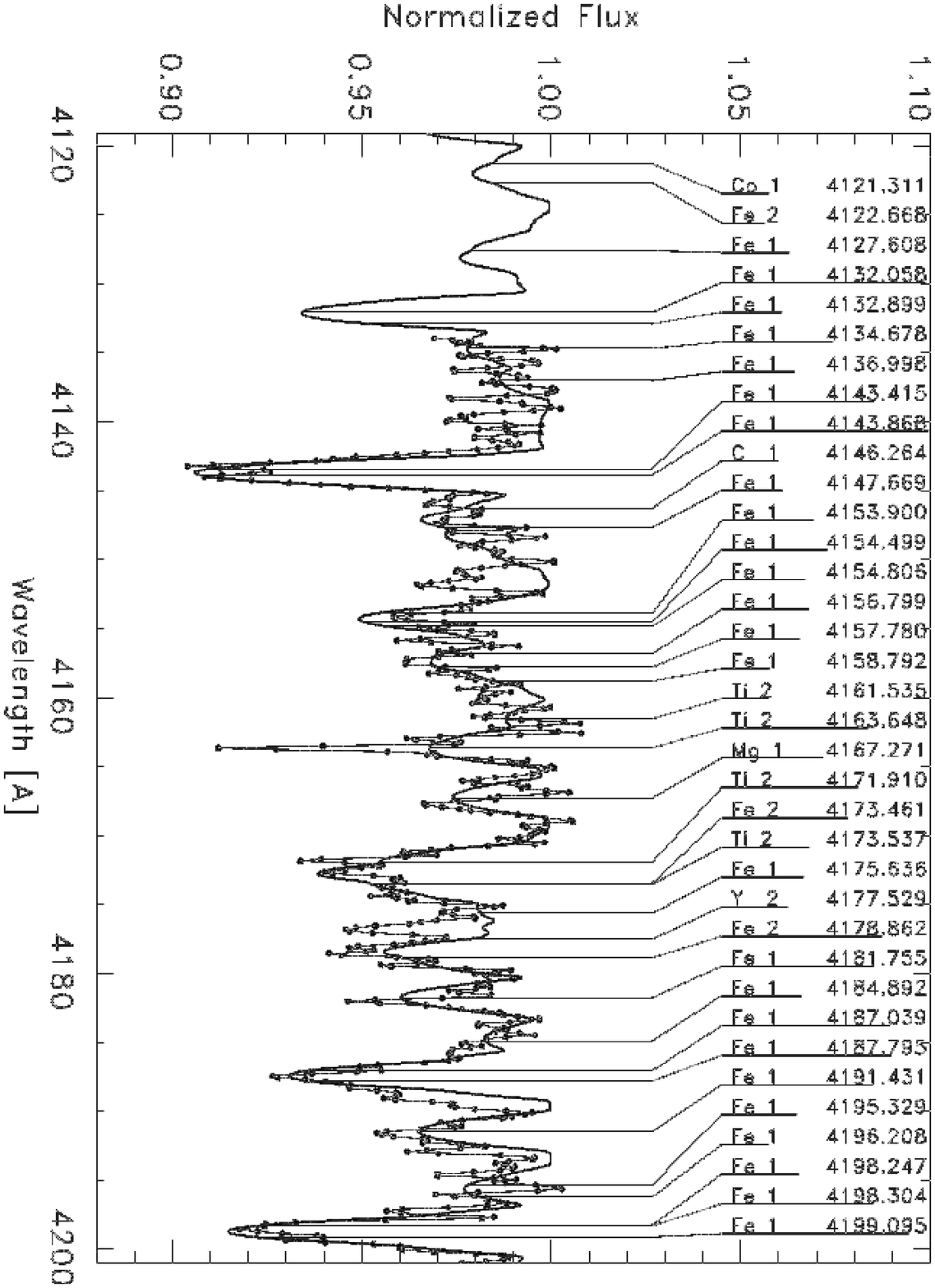}
                      \includegraphics[angle=90]{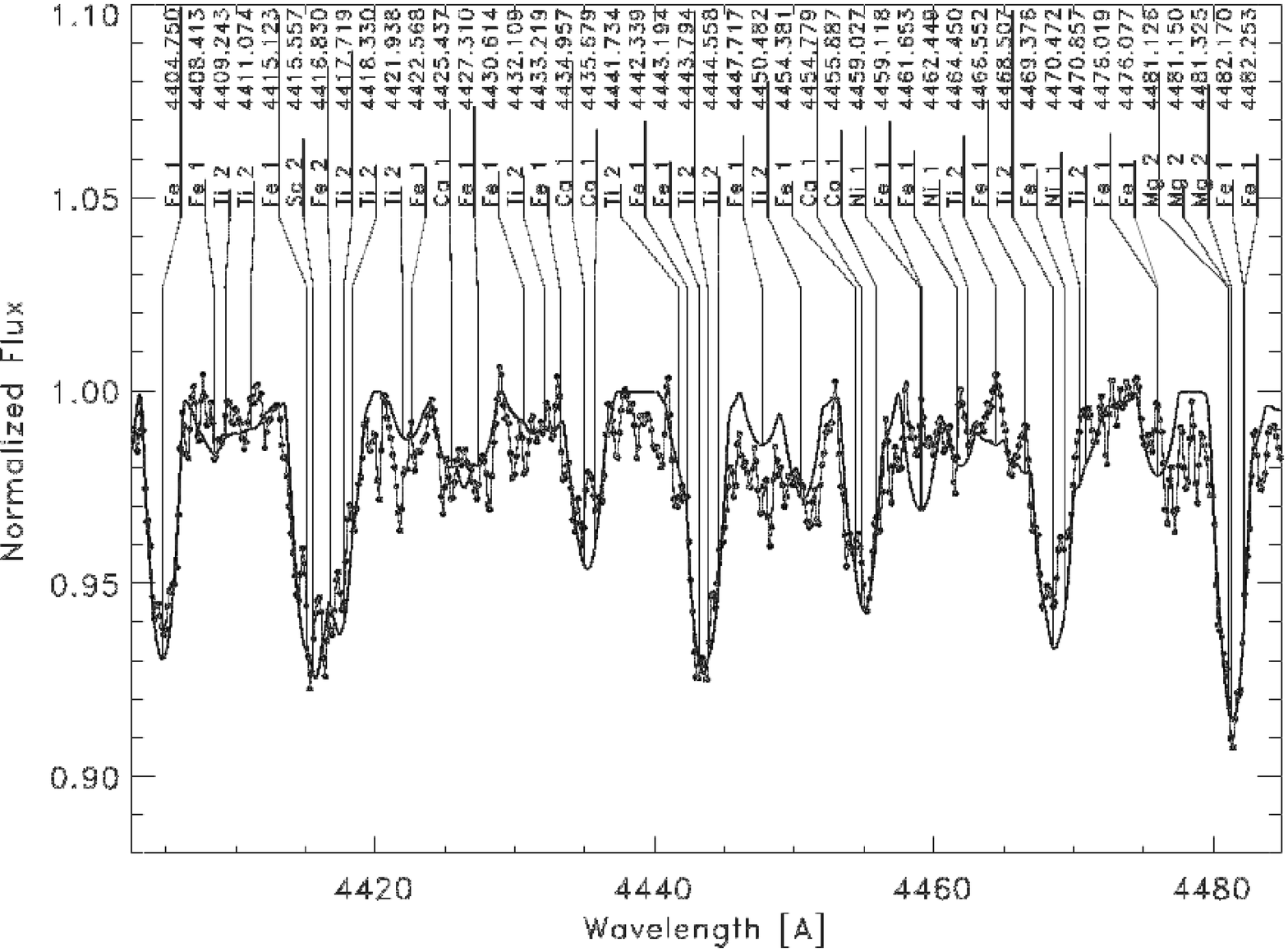}}
  \caption{Two sections of the spectra of HD\,106223 (upper plots, observations
   obtained at Asiago) and HD\,142703 (lower plots, observations obtained at 
   LNA). The characteristics of the different observations are given in Table~\ref{obs_tab}.
  Observations: dots connected with thin lines; Calculations with best-fit 
  parameters and abundances (Tables~\ref{param_tab} and \ref{abun_tab_1}):
  thick lines. Lines with unconvolved central linedepths larger than 16 and
  10~\% in the left and right wavelength regions, respectively,
  are labelled.}
  \label{spectra}
\end{figure*}

High resolution and classification resolution spectra with high signal to noise 
have been obtained at various sites as listed in Table~\ref{obs_tab}. 
Fig.~\ref{spectra} shows a part of the high resolution spectra of two stars with the most similar
atmospheric parameters (see Sect.~3) obtained with two different instruments.
They have been reduced with the 
{\tt ccdred} and {\tt echelle} packages of NOAO IRAF following \citet{Will:94}.

\begin{figure}
\resizebox{\hsize}{!}{\includegraphics[angle=90]{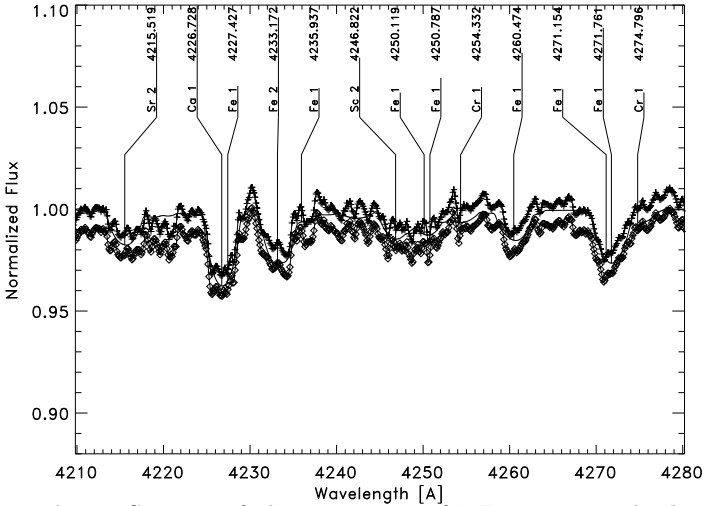}}
  \caption{Section of the spectrum of HD\,110411 with the continuum
  of the observed spectrum raised (pluses) and lowered (diamonds) by 0.5\%.
  This seems to be the maximum deviation from the ``true'' continuum as a
  comparison with the calculated spectrum (solid line) shows. 
  For better visibility, the observations 
  have been smoothed with a boxcar average using a width of 5 pixels.
  The flux scale is the same as in the other figures.}
  \label{continuum}
\end{figure}

Particular care has been devoted to the normalization of the spectra to the
continuum, which is in general rather difficult for broad lined stars.
But in the case of metal poor stars as examined here, the problem is not
as serious, because their spectra show sufficiently few lines, so that there
are enough continuum sections in between for every echelle order.
The interactive curve fitting tool, ICFIT, within IRAF is a convenient help in performing this task.
A cubic spline is fit to the data points contained in one echelle order,
iteratively excluding those points from the fit whose residuals are greater than 
the noise level of the observations. 
The normalized observed spectra were compared to
synthetic spectra to check if the chosen continuum sections are
devoid of spectral lines. Our experience with VALD based on analyses of
stars with a wide range of metallicities justifies the assumption that
these sections indeed do not contain any unknown significant spectral lines.

The error introduced by deviations of the fit from the real continuum level 
influences mostly the abundances of elements for which only few lines can be used. 
When using many lines distributed over a large
wavelength range for the abundance determination of an element,
the continuum placement error is assumed to be negligible compared to the
scatter of the ``line'' abundances (see below) from which
the abundance error has been determined.
The spectrum of HD\,110411 (one of
the broadest lined stars) has been raised and lowered by 0.5\%, which seems
to be a maximum for an incorrect continuum fit (see Fig.~\ref{continuum}). 
The errors resulting from such a displaced continuum amount 
to 0.1~dex for C, Ca and Ti, and 0.2~dex for Mg and Sr for HD\,110411. 
They have been obtained by dividing the abundance differences of several elements relative
to our final values by the square root of the number of lines 
contained in separate orders.

The procedure of the abundance analysis which has been established in the
Vienna working group was described in detail for example by
\citet{Gelb:97} and \citet{Heit:98}, and is summarized in \citet{Heit:00}. 
Model atmospheres have been computed with the ATLAS9 program \citep{Kuru:93a}, but using
the CM model \citep{Canu:91} for the treatment of convection. 
The Rosseland opacities and the opacity distribution functions (ODFs) 
have been taken from \citet{Kuru:93a,Kuru:93b}.
For a discussion of different convection models and their usefulness 
for the abundance analysis of \LB\ stars as well as ODFs calculated 
for the \LB\ abundance pattern see \citet{Heit:98}.
Synthetic spectra have been calculated with the programs SYNTH and
ROTATE \citep{Pisk:92}, using log~$gf$ values and further atomic data 
extracted from the Vienna atomic line database (VALD), which
contains the most recent laboratory data \citep{Kupk:99},
and new calculations of van der Waals damping constants \citep{Bark:00a}.

The physics used in the SYNTH program are the same as that described in \citet{Vale:96}.
It includes sophisticated H-line treatment \citep{Bark:00b} and it uses an adaptive
wavelength grid. A satisfactory comparison of the SYNTH program with two other spectrum
synthesis programs can be found in \citet{Ilie:96}.
An additional argument for the
SYNTH program being consistent with the literature is given by the successful
reproduction of Vega abundances by \citet{Kupk:96a}.\\

The starting values for the effective temperature (\Teff) and the surface
gravity (\logg) were derived 
from Str\"omgren photometry using the program of 
\citet{Napi:93}. These have been varied after the calculation
of a set of Fe ``line'' abundances, which were obtained
by fitting the observed and calculated line profiles individually
for each line.
The best-fit \Teff, \logg\ and microturbulence velocity (\Vmicro) were found
when ``line'' abundances did not correlate with 
excitation energy, ionisation stage and equivalent width,
respectively. In addition, the line abundances have been tested for
a dependence on the effective Land\'e factor, 
which indicated the absence of large magnetic fields in the
program stars, which is consistent with the results of \citet{Bohl:90}.

The internal accuracy of individual line abundances was estimated
by fitting synthetic line profiles to the observations, but putting
more weight on the maximum noise peaks and in a second step on the minima.
The difference ($\sigma_{\rm l}$) of the such determined line abundances is a measure for
the accuracy which can be achieved with the observed spectra for individual lines.
The abundances and standard errors for each element have been obtained by averaging
the ``line'' abundances for each program star weighted by 1/$\sigma_{\rm l}^2$.

All the program stars have \Vsini\ values near 100\,\kms
or higher. Because only few unblended lines can be found in the
spectra, we had to use blends consisting of several lines from
one element. Therefore it was necessary to estimate the contribution 
of individual blend components and their sensitivity to abundance changes.
Some blends of iron lines with equal ionisation stage
considered for the analysis of HD\,106223 are listed in Table~\ref{equw_change}.
The relative difference in equivalent width of each blend component
has been calculated for synthetic spectra with an iron abundance difference
of 0.4~dex. Blends have been regarded as useful if the contribution to the 
equivalent width change of all but one component was less than 50\% 
or if two components equally contributing to equivalent width changes
 had similar atomic parameters (equal
sensitivity to \Teff, \logg\ and \Vmicro). The atomic data for these,
a blend dominating components, have been used in the trend analysis.

\begin{table}
\caption{Examples for the relative change of equivalent width ($\Delta W$) 
of iron lines when changing the abundance by 0.4~dex.
Also given are the wavelength $\lambda$, the ionisation stage,
the lower excitation potential $E_{\rm low}$ and the effective Land\'e factor
$g_{\rm eff}$. 
Blends where one or two lines are marked with a bullet have been selected for 
analysis, using the atomic parameters of the marked lines for the atmospheric
parameter tests.}
\label{equw_change}
\begin{tabular}{rccrr}
\hline\hline
 $\lambda$ [\AA]  & Ion & $E_{\rm low}$ [eV] & $g_{\rm eff}$ & $\Delta W$ [\%] \\
\hline
 4062.441  &  1 &  2.845 & 1.69 & 42 \\
 4063.276  &  1 &  3.368 & 1.15 & 25 \\
 4063.594  &  1 &  1.557 & 1.09 & 30 \\
 4063.627  &  1 &  4.103 & 1.44 &  0 \\
 4064.450  &  1 &  1.557 & 0.34 &  3 \\
\hline
 4135.271  &  1 &  3.397 & 1.59 &  5 \\
 4135.755  &  1 &  4.191 & 1.50 &  2 \\
 4136.521  &  1 &  3.368 & 1.35 & 12 \\
 $\bullet$4136.998  &  1 &  3.415 & 1.13 & 73 \\
 4137.420  &  1 &  4.283 & 1.24 &  8 \\
\hline
 $\bullet$4187.039  &  1 &  2.449 & 1.48 & 47 \\
 4187.587  &  1 &  3.430 & 1.33 &  7 \\
 4187.617  &  1 &  3.640 & 1.10 &  0 \\
 $\bullet$4187.795  &  1 &  2.425 & 1.47 & 46 \\
\hline
 4582.835  &  2 &  2.844 & 1.63 & 31 \\
 4582.940  &  1 &  2.845 & 1.44 &  0 \\
 $\bullet$4583.837  &  2 &  2.807 & 1.14 & 63 \\
 4583.999  &  2 &  2.704 & 2.06 &  4 \\
 4584.715  &  1 &  3.603 & 1.66 &  1 \\
\hline\hline
\end{tabular}
\end{table}


\section{Program stars and results}

\begin{table}
\caption{Program stars and their characteristics, excluding the spectroscopic binary systems. Columns 3 and 4 indicate if the depression centered at 1600\AA\ is present and if the C/Al equivalent width ratio measured from IUE spectra is enhanced, as is typical for \LB\ stars \citep{Fara:90,Sola:98}, and column 5 shows if the IR spectral properties support the \LB\ classification \citep{Andr:95}.}
\label{stars}
\begin{tabular}{rrccc}
\hline\hline
HD     & Spectral type                  & 1600\AA & C/Al & IR \\
\hline
74873 & kA0.5hA5mA0.5~V\,$\lambda$\,Boo$^1$ &     &     &     \\
101108 &      A3~IV-V\,($\lambda$\,Boo)$^2$ & yes & no  & yes \\
106223 &    kA3hF3mA3~V\,$\lambda$\,Boo$^2$ &     &     & yes \\
107233 &    kA2hF1mA2~V\,$\lambda$\,Boo$^2$ &     &     &     \\
110411 &        A0~Va\,($\lambda$\,Boo)$^3$ & yes & yes & yes \\
142703 &   kA1hF0mA1~Va\,$\lambda$\,Boo$^4$ &     & yes &     \\
168740 &       hA7mA2~V\,$\lambda$\,Boo$^2$ &     &     &     \\
170680 &       A0~Van\,($\lambda$\,Boo)$^5$ &     & yes &     \\
\hline\hline
\multicolumn{5}{l}{} \\
\multicolumn{5}{l}{$^1$ \citet{Paun:97b}, $^2$ \citet{Paun:00},} \\
\multicolumn{5}{l}{$^3$ \citet{Gray:88}, $^4$ \citet{Gray:91}, $^5$ \citet{Gray:87}} \\
\end{tabular}
\end{table}

The program stars examined in this work, except for the two spectroscopic
binary systems HD\,84948 and HD\,171948, are listed in Table~\ref{stars}
along with their characteristics found in the literature.
Calculated spectra of two stars are shown in Fig.~\ref{spectra}
with the corresponding observations.
The final parameters used for the analyses are listed in Table~\ref{param_tab}.

These spectroscopically determined atmospheric parameters 
do not differ significantly from the
photometrically derived ones except for HD\,106223 and HD\,107233, for which
higher and lower \logg\ values, respectively, resulted in more consistent
line abundances. 
The best-fit atmospheric parameters are uncertain by up to 400~K for \Teff,
0.4 for \logg\ and 0.5~\kms\ for \Vmicro, which has been estimated
from the standard error of the gradient of the linear fit between the atomic
parameters determining the sensitivity of the lines to the respective atmospheric
parameters (see Sect.~2).
The error of \Vsini\ has been estimated from line profile fits to be 10~\kms.

\begin{figure*}
\resizebox{\hsize}{!}{\includegraphics[angle=270]{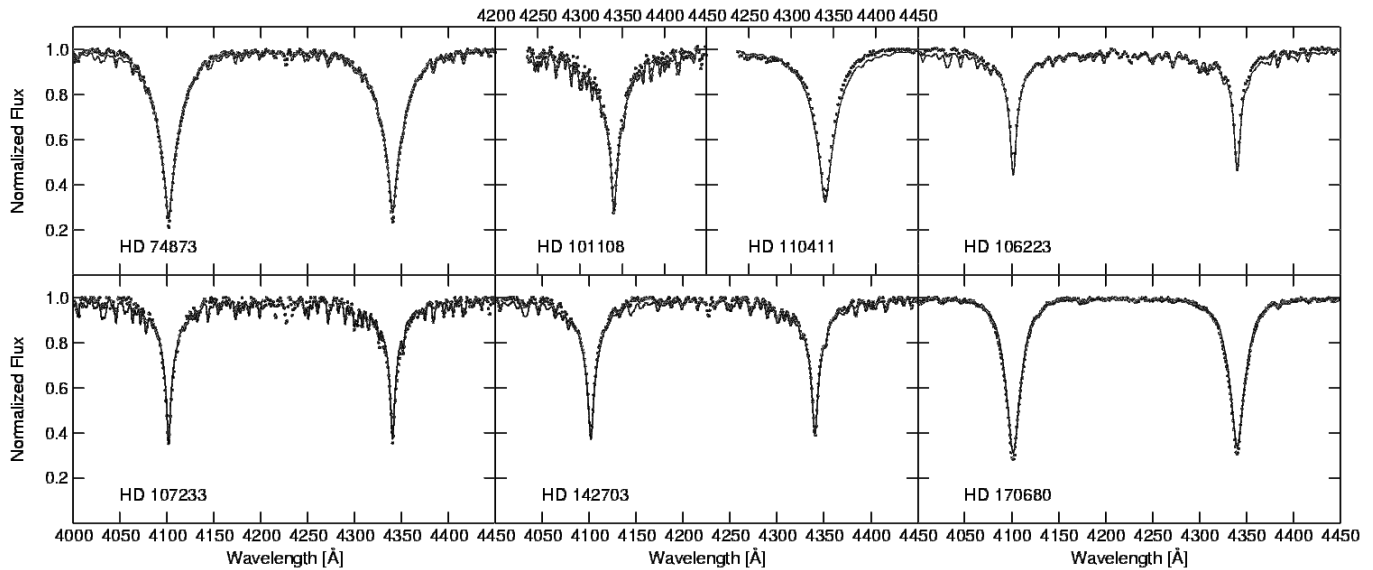}}
  \caption{Classification resolution hydrogen line profiles of the program stars.
  Dots: Observations; Lines: Calculations with parameters as in 
  Table~\ref{param_tab}, except for HD\,74873 (\logg=4.3) and HD\,106223 
  (\Teff=6800~K). The characteristics of the observations are given in Table~\ref{obs_tab}.}
  \label{hydrogen}
\end{figure*}

An additional check of the parameters can be obtained from 
H$_{\beta}$, H$_{\gamma}$ and H$_{\delta}$ lines, which
unfortunately are distributed over several orders in the echelle spectra.
This fact and their broad wings makes it very difficult to normalize 
these regions to the continuum.
Fortunately, classification resolution spectra were available
for most of the investigated stars \citep{Paun:00}, allowing to
estimate the continuum flux level for H$_{\gamma}$ and H$_{\delta}$ lines
fairly well (Fig.~\ref{hydrogen}).

The mean element abundances of all program stars are given in 
Table~\ref{abun_tab_1}.
Formal errors calculated from the line abundances are given in parantheses
in units of the last significant digit and are representative for the quality
of the observational data. Other error sources have to be considered additionally,
like errors in \Teff\ and \logg, which would change the element abundances, but
increase the scatter in the line abundances for the individual elements.
These changes depend largely on the chosen model atmosphere parameters and not
on the peculiarity of the investigated stars.
Values which are typical for the parameter space we are investigating can be
found for the \LB\ star $\pi^1$~Ori (\Teff=8750~K, \logg=4.0)
in \citet{Venn:90} and are corroborated by our own experience. 
We have repeated the abundance analysis for HD\,168740 focussing on Fe,
which is most suitable for this sort of test because of the relatively large number
of unblended lines with well-known atomic parameters in two ionization stages.
Decreasing the effective temperature by 600~K decreases the Fe abundance by 0.3~dex,
but increases the rms deviation of the line abundances from 0.15 to 0.24.
Increasing the \logg\ value by 0.6~dex increases the Fe abundance
only by 0.02~dex but increases the rms deviation of the line abundances to 0.19.
Uncertainties arising from the continuum fit have been discussed in Sect.~2.
The abundances of the neutral and singly ionized elements agree within the
observational error limits, indicating a properly chosen atmosphere.

For HD\,84948 and HD\,171948 the results of the analysis of \citet{Paun:98b}
have been confirmed and extended by using the IDL program BROTATE written 
by O.P. Kochukhov (private communication).
Within the error limits the two components of HD\,171948 have the same abundances 
and are therefore listed only once in Table~5b.

\begin{table}
\caption{Best-fit atmospheric parameters of the program stars. 
\logZ\ refers to the abundances scale of the ODFs.}
\label{param_tab}
\begin{tabular}{rrlccc}
\hline\hline
         & \Teff & \logg & \Vmicro & \logZ & \Vsini \\
 HD      & [K]   & [log             & [\kms] & & [\kms] \\
         &       &     cm~s$^{-2}$] &        & &        \\
\hline
 74873   &  8900 & 4.6 & 3.0 & $-$1.0 & 130 \\ 
 84948A$^{\ast}$  &  6600 & 3.3 & 3.5 & $-$1.0 &  45 \\
 84948B$^{\ast}$  &  6800 & 3.5 & 3.5 & $-$1.0 &  55 \\
101108   &  7900 & 4.1 & 3.0 & $-$1.0 &  90 \\
106223   &  7000 & 4.3 & 3.0 & $-$1.5 & 100 \\
107233   &  6800 & 3.5 & 3.0 & $-$1.5 &  80 \\
110411   &  9200 & 4.6 & 3.0 & $-$1.0 & 150 \\
142703   &  7000 & 3.7 & 3.0 & $-$1.5 & 100 \\
168740   &  7900 & 3.5 & 3.0 & $-$1.0 & 130 \\
170680   & 10000 & 4.1 & 2.0 & $-$0.5 & 200 \\
171948$^{\ast}$  & 9000 & 4.0 & 2.0 & $-$1.5 & 15/10$^{\ast\ast}$ \\
\hline\hline
\multicolumn{6}{l}{} \\
\multicolumn{6}{l}{$^{\ast}$ from Paunzen et al. (1998), $^{\ast\ast}$ A/B} \\
\end{tabular}
\end{table}

In the following, comments on the results of the individual stars are given.

\noindent{\em HD\,74873:}
The projected rotational velocity of 130~\kms\ derived from the high resolution
spectra corresponds to the similarly high values found for many other
\LB\ stars. It is somewhat higher than the value given in the catalogue
of \citet{Uesu:82} and contradicts the value of 10~\kms\
given by \citet{AbtM:95}. Thus the star has to be 
rejected from the (small) list of possibly slowly rotating \LB\ stars.
Abundances for 12 elements could be determined. The underabundances
of the heavier elements lie in the range of $-$0.5 to $-$1 dex, except
for Ni, which seems to be almost solar abundant. 
The three lighter elements show solar (O) or +0.5 dex higher (C, Na) abundances.

\noindent{\em HD\,84948:}
For this spectroscopic binary system, abundances for 5 additional
heavy elements have been determined, using the same parameters as
\citet{Paun:98b}. The two stars belong to the small number
of \LB\ stars for which Zn abundances could be derived. At least for HD\,84948~A,
Zn is as deficient as the other heavy elements. This is contradictory to the prediction of
accretion theory that Zn should resemble the abundances of the light elements
(C, N, O, S) because of its similarly low condensation temperature \citep{vanW:92}.
The same has been found already for HD\,84123 \citep{Heit:98}.
Note that the abundance of Y is the highest found for all heavy elements
in both stars.

\noindent{\em HD\,101108:}
Abundances for 14 elements could be determined for this star,
as well as an upper limit for the abundance of N.
The resulting abundance pattern is typical for \LB\ stars,
with a solar C abundance and a wide range of underabundances (up to $-$1~dex) 
for the heavy elements. As in the previous two stars, Y shows the highest 
abundance of all heavy elements, namely a solar one.
The abundance ratio of Al to C seems to contradict the results of the 
UV measurements presented in Table~\ref{stars}, 
but as the Al abundance is based on only one line,
its value remains questionable.

\noindent{\em HD\,106223:}
Element abundances for HD\,106223 have been derived already 40 years ago
by \citet{Burb:56} within their examination
of ``Five Stars which Show some of the Characteristics of Population~{\sc ii}''.
A direct comparison to the results of the present work is not possible, because 
of the different method (curve of growth analysis) and effective temperatures
used by them. Furthermore, the abundances are listed relative to
95\,Leo, which most probably has non-solar composition \citep{Rens:91,AbtM:95}. 
However, the abundances they derived for HD\,106223 are lower than that of
HD\,84123 \citep{Heit:98}, 
but higher than that of $\lambda$\,Boo \citep{Venn:90}, which is consistent
with the results of the present work. Interestingly,
four of the five stars included in the above mentioned work have been assigned
to the \LB\ group later on. 
The 15 element abundances determined for HD\,106223 are characteristic for 
\LB\ stars -- a slightly higher than solar C abundance and a $-$1.5~dex mean 
underabundance of heavy elements.
Zn shows a higher than average abundance but not a solar one,
and the Y abundance is remarkably high again, although it is based on only 
one spectral line.

\noindent{\em HD\,107233:}
The rotational velocity of this star has been determined for the first
time in the present work and turned out to have a value typical for 
\LB\ stars (\Vsini\ = 80~\kms). 
The derived abundances confirm the classification of HD\,107233 as a
\LB\ star. C is only slightly deficient whereas the heavy elements are
on the average underabundant by $-$1.4~dex, with an enhancement
of Mg and Y and a deficiency of Al relative to Fe.

\begin{table*}
\caption{Abundances of the program stars. For each element $X$ the value 
$\log\,\left(\frac{N_X}{N_{\rm tot}}\right) - 
\log\,\left(\frac{N_X}{N_{\rm tot}}\right)_{\odot}$ is given,
where the solar abundances are taken from \citet{Grev:96}.
The columns containing the numbers of lines per element for two ionisation
stages are labeled N. 
Errors in parentheses are standard errors of the mean line abundances and are 
given in units of the last significant digit. No errors are given when only one
line has been used.}
\label{abun_tab_1}
\begin{tabular}{cclclclclcl}
\hline\hline
Element & N & 74873   & N      & 84948\,A  & N    & 84948\,B  & N    & 101108    & N    &  106223   \\
\hline
C  & 8/0  & $+$0.6(1) &        &           &      &           & 7/0  & $+$0.1(1) & 6/0  &  +0.3(1)  \\
N  &      &           &        &           &      &           & 1/0  & $<+$0.7   &      &           \\
O  & 1/0  &   +0.0    &        &           &      &           &      &           &      &           \\
Na & 1/0  & $+$0.5    &  2/0   & $-$0.3(4) & 2/0  &   +0.0(3) &      &           &      &           \\
Mg & 4/1  & $-$0.9(1) &  5/1   & $-$1.2(5) & 7/1  & $-$1.0(4) & 6/1  & $-$0.3(1) & 2/1  & $-$1.7(1) \\
Al & 1/0  & $-$1.1    &        &           &      &           & 1/0  & $-$0.9    & 1/0  & $-$2.1    \\
Si &      &           &  0/2   & $-$0.8(4) & 0/2  & $-$0.6(5) & 1/1  & $-$0.5(3) & 1/0  & $-$1.7    \\
Ca & 1/1  & $-$0.7(1) &  11/0  & $-$1.3(5) & 9/0  & $-$0.8(4) & 7/0  & $-$0.6(1) & 4/1  & $-$1.9(1) \\
Sc & 0/1  & $-$0.4    &  0/6   & $-$1.4(4) & 0/6  & $-$0.7(5) & 0/3  & $-$0.6(2) & 0/2  & $-$2.1(3) \\
Ti & 0/6  & $-$0.9(1) &  1/11  & $-$1.4(3) & 1/20 & $-$0.6(4) & 0/17 & $-$0.3(1) & 0/11 & $-$1.6(1) \\
Cr & 2/1  & $-$0.3(2) &  7/6   & $-$1.1(4) & 4/6  & $-$0.9(5) & 4/7  & $-$0.6(1) & 4/2  & $-$1.7(1) \\
Mn &      &           &        &           &      &           & 6/0  & $-$0.4(1) & 5/0  & $-$1.7(2) \\
Fe & 14/8 & $-$0.6(1) &  43/11 & $-$1.1(3) & 47/9 & $-$0.9(2) &36/10 & $-$0.60(4) &35/11 & $-$1.44(4) \\
Ni & 1/0  & $-$0.2    &  4/0   & $-$0.6(4) & 4/0  & $-$0.6(4) & 6/0  & $-$0.2(1) & 5/0  & $-$1.1(1) \\
Zn & 1/0  & $<+$0.9   &  2/0   & $-$0.9(4) & 1/0  & $-$0.4    &      &           & 2/0  & $-$0.8(1) \\
Sr & 0/2  & $-$0.9(4) &  0/2   & $-$1.3(5) & 0/2  & $-$0.8(4) & 0/2  & $-$0.7(2) & 0/1  & $-$1.7    \\
Y  &      &           &  0/3   & $-$0.5(4) & 0/2  & $-$0.2(4) & 0/3  & $+$0.0(2) & 0/1  & $-$0.5    \\
Ba &      &           &  0/4   & $-$1.0(4) & 0/2  & $-$0.8(5) & 0/2  & $-$0.7(2) & 0/2  & $-$1.6(1) \\
\hline\hline
\end{tabular}
\end{table*}

\noindent{\em HD\,110411:}
This star has already been analysed
by \citet{Stue:93}, who gives abundances for only three elements 
(Ca, Fe and Sr), which are all equally underabundant by $-$1.0~dex. This
result compares well to the new abundance determinations ($-$0.7,
$-$1.0 and $-$1.2~dex, respectively), if the errors and small differences
in the atmospheric parameters (St\"urenburg
used 8970~K, 4.36 and 175~\kms\ for \Teff, \logg\ and \Vsini, respectively) 
are taken into account.  
Abundances for 6 additional elements are given here, 
because a much larger wavelength range could be examined in the present work.
Of these, the C abundance of +0.1~dex and the (admittedly uncertain) O abundance of +0.4~dex
are of particular interest, confirming the \LB\ character previously 
attributed to this star due to the properties of its UV spectrum.
Further attention should be given to Mg and Si, which seem to be
less deficient than the other heavy elements.

\addtocounter{table}{-1}
\begin{table*}
\caption{continued.}
\label{abun_tab_2}
\begin{tabular}{cclclclclclcl}
\hline\hline
Element & N &  107233 & N    & 110411    & N      & 142703    & N     & 168740    & N    & 170680    & N    & 171948A/B \\
\hline
C  & 2/0  & $-$0.3(3) & 6/0  & $+$0.1(2) & 6/0    & $-$0.1(1) &       &           &      &           & 3/0  & $<-$0.5   \\
N  & 1/0  & $<+$0.2   &      &           &        &           &       &           &      &           &      &           \\
O  &      &           & 1/0  & $+$0.4    &        &           &       &           &      &           & 2/0  & $-$0.6(4) \\
Na &      &           &      &           &        &           &       &           &      &           & 5/0  & $+$0.2(4) \\
Mg & 2/1  & $-$0.8(1) & 3/1  & $-$0.5(2) & 1/1    & $-$1.2(1) & 1/1   & $-$0.9(1) & 0/2  & $-$0.2(1) & 0/1  & $-$2.4    \\
Al & 1/0  & $-$2.2    &      &           &        &           &       &           &      &           &      &           \\
Si &      &           & 0/1  & $-$0.3    &        &           &       &           &      &           & 0/2  & $-$1.6(4) \\
S  &      &           &      &           & 1/0    & $<-$0.7   &       &           &      &           &      &           \\
Ca & 4/0  & $-$1.4(1) & 1/1  & $-$0.7(1) & 4/0    & $-$1.2(1) & 4/0   & $-$0.6(1) &      &           &      &           \\
Sc & 0/1  & $-$1.7    & 0/1  & $-$1.1    & 0/1    & $-$1.5    & 0/2   & $-$1.1(4) &      &           &      &           \\
Ti & 0/11 & $-$1.4(1) & 0/6  & $-$0.9(1) & 0/9    & $-$1.4(1) & 0/10  & $-$1.1(1) & 0/5  & $-$0.5(1) & 0/5  & $<-$3.0    \\
Cr & 2/1  & $-$1.5(2) &      &           & 2/3    & $-$1.39(4) & 3/6   & $-$1.1(1) & 0/3  & $-$0.4(2) & 0/2  & $-$1.8(5) \\
Mn & 4/0  & $-$1.3(1) &      &           &        &           &       &           &      &           &      &           \\
Fe & 27/5 & $-$1.40(3) & 10/5 & $-$1.0(1) & 21/8  & $-$1.52(3) & 16/11 & $-$0.89(3) & 1/9  & $-$0.4(1) & 5/10 & $-$1.6(4) \\
Co & 1/0  & $-$1.0    &      &           &        &           &       &           &      &           &      &           \\
Ni &      &           &      &           & 3/0    & $-$0.9(2) &       &           &      &           &      &           \\
Zn &      &           &      &           & 2/0    & $<-$0.6   &       &           &      &           &      &           \\
Sr & 0/1  & $-$1.3    & 0/2  & $-$1.2(2) & 0/1    & $-$1.2    & 0/1   & $-$1.0    &      &           &      &           \\
Y  & 0/2  & $-$0.9(2) &      &           &        &           &       &           &      &           &      &           \\
Zr & 0/1  & $-$1.5    &      &           &        &           &       &           &      &           &      &           \\
Ba &      &           &      &           & 0/1    & $-$1.5    & 0/1   & $-$0.5    &      &           &      &           \\
\hline\hline
\end{tabular}
\end{table*}

\noindent{\em HD\,142703:}
The C abundance determined in the present work can be compared with the analysis
of \citet{Paun:99a}, which is based on near-IR spectra and
non-LTE calculations. Their work resulted in a C abundance of $-$0.5~dex,
which is 0.4~dex lower than the abundance obtained here.
This difference can only in part be attributed to the lower \Vsini\ of 75~\kms\ 
used by \citet{Paun:99a}.
As another source for the discrepancy, the use of different sets of atomic
data was suspected, because the analysis for this star was made at earlier
times than the near-IR analysis, with the then available VALD-1 data
\citep{Pisk:95}. But a comparison of the old and new atomic data for the
lines used showed that the differences are very small and would cause
uniformly distributed positive and negative shifts in the line abundances.
This provides additional indication for an inconsistency of abundances derived from
optical and near-IR spectra, as has already been realized by
\citet{Paun:99a} from a comparison of their results with that of \citet{Stue:93}. 
The Mg, Cr, Fe and Sr abundances agree well with the
results of \citet{Nort:94a,Nort:94b}, if only the spectral lines used in common are regarded.
The abundance pattern of HD\,142703 is consistent with its
\LB\ classification with underabundances of heavy elements ranging
from $-$1 to $-$1.5~dex. The \LB\ character is confirmed by
the O abundance of $-$0.1~dex \citep{Paun:99a}.
An estimation for the upper limits of the S and Zn abundances 
has been made, which indicates that these elements are not solar abundant.

\noindent{\em HD\,168740:}
The Mg, Cr, Fe and Sr abundances determined from the common spectral
lines are again similar to that of \citet{Nort:94a,Nort:94b},
although they used a lower \Vsini\ of 90~\kms. In total, abundances 
could be determined for eight heavy elements, resulting in
underabundances around $-$1~dex, with Ca and Ba showing about
0.5~dex higher than average abundances. Together with the C and O abundances 
given by \citet{Paun:99a} as $-$0.4 and +0.0~dex, respectively,
HD\,168740 exhibits a typical \LB\ abundance pattern.

\noindent{\em HD\,170680:}
HD\,170680 is the hottest star analysed so far, and it is the fastest
rotator of the sample studied here. The \Vsini\ of 200~\kms\ 
derived during the analysis confirms the value obtained by 
\citet{AbtM:95}, as opposed to earlier determinations of 305~\kms\ 
\citep{Uesu:82}. These two properties make an abundance
analysis very difficult, because only few suitable lines are seen in the
spectrum. On the other hand, the heavy elements are only moderately
deficient in this star (up to $-$0.5~dex) and therefore it was possible 
to determine rather accurate abundances for four elements. Together with the 
solar C and O abundances \citep{Paun:99a} they form a mild \LB\ abundance pattern.

\noindent{\em HD\,171948:}
The 1997 observations obtained for this spectroscopic binary (SB) system
had a lower resolution than the previous observations which lead to the
discovery of the binarity. Therefore, in the new spectra, 
the corresponding lines of the two components were combined in blends,
leaving about half of each profile distinguishable. 
The known \Vsini\ values
(in particular their being not equal) and atmospheric parameters \citep{Paun:98b}
made it possible to determine the radial velocity difference of the two SB 
components in the new observations, which was only half the value of 1996.
Consequently, the observed spectra could be reproduced by synthetic ones,
using the parameters and abundances determined previously.
Additionally, upper limits and estimates for abundances of other
elements could be derived, in particular for some light elements.
Lower limits for the Na abundance have been determined by 
the fit to the resonance lines $\lambda\lambda~5890/6$, which yielded
LTE abundances of +1.4 and +0.7~dex, respectively. The NLTE effect 
can amount to $-$0.8~dex for these lines \citep{Mash:96}.
The combination with upper limits given by several very weak lines
($\lambda\lambda$~4978.5, $\lambda\lambda$~4982.8, $\lambda\lambda$~6160.7),
results in a solar-like Na abundance, whereas C is deficient by at least
0.5~dex. Oxygen shows an underabundance of $-$0.6~dex, which is still
high compared to the heavy element abundances.
These abundance values are the same for both components within the
error limits.

\begin{figure*}
\resizebox{12cm}{!}{\includegraphics{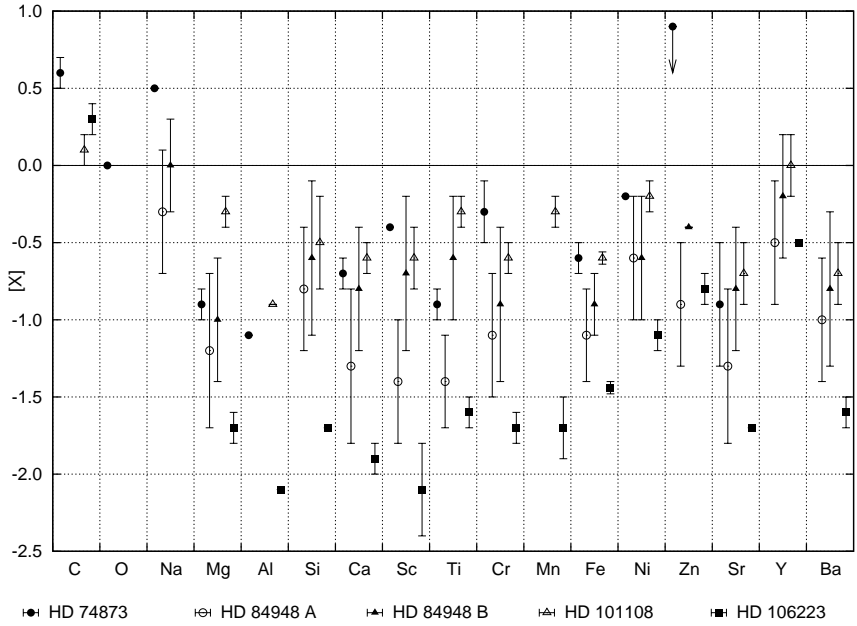}}
  \hfill
  \parbox[b]{55mm}{
  \caption{Abundances of the program stars from Table~\ref{abun_tab_1} (first five stars).
           Upper limits are marked by arrows pointing downwards.}
  \label{abun_fig_1}}
\end{figure*}

\begin{figure*}
\resizebox{12cm}{!}{\includegraphics{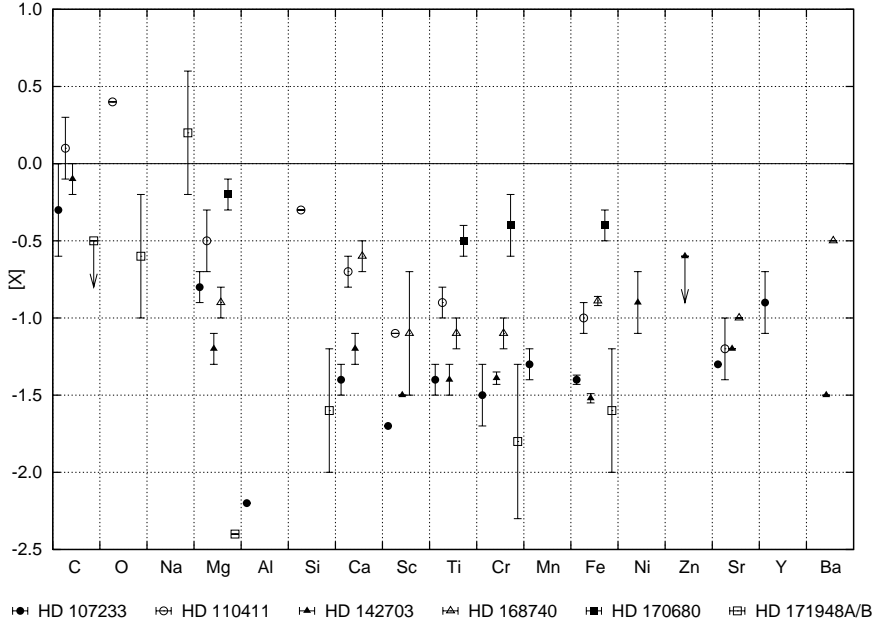}}
  \hfill
  \parbox[b]{55mm}{
  \caption{Abundances of the program stars from Table~\ref{abun_tab_2} (last six stars).}
  \label{abun_fig_2}}
\end{figure*}

The abundance patterns of all program stars are illustrated in 
Figs.~\ref{abun_fig_1} and \ref{abun_fig_2}.
These Figures, as well as Table~\ref{abun_tab_1}, include
also the abundances of the spectroscopic binaries already reported
in \citet{Paun:98b}.

\section{The \LB\ star abundance pattern}
\label{pattern}

\begin{table*}
\caption{References for element abundances determined for \LB\ stars. A: \citet{Adel:99}, Ch: \citet{Cher:98}, H: \citet{Heit:98} and this work, K: \citet{Kamp:01}, M: \citet{Mart:98}, N: \citet{Nort:94a,Nort:94b}, P: \citet{Paun:99b}, So: \citet{Sola:01}, St: \citet{Stue:93}, V: \citet{Venn:90}. The HD numbers of spectroscopic binaries treated as single stars in the abundance analyses are given in parantheses.}
\label{ref_tab}
\begin{tabular}{rp{1.1em}p{1.1em}p{1.1em}p{1.1em}p{1.1em}p{1.1em}p{1.1em}p{1.1em}p{1.1em}p{1.1em}p{1.1em}p{1.1em}p{1.1em}p{1.1em}p{1.1em}p{1.1em}p{1.1em}p{1.1em}p{1.1em}p{1.1em}p{1.1em}}
\hline\hline
HD	& C	& N	& O	& Na	& Mg	& Al	& Si	& S	& Ca	& Sc	& Ti	& Cr	& Mn	& Fe	& Co	& Ni	& Zn	& Sr	& Y	& Zr	& Ba	\\
\hline
319	& St	&	&	& St	& St	& St	& St	&	& St	& St	& St	& St	&	& St	&	&	&	& St	&	&	& St	\\
11413	& St	&	&	&	& St	&	& St	&	& St	& St	& St	& St	&	& St	&	&	&	& St	&	&	& St	\\
15165	& Ch	&	& Ch	& Ch	& Ch	&	& Ch	& Ch	& Ch	& Ch	& Ch	& Ch	&	& Ch	&	& Ch	&	& 	&	&	& Ch	\\
31295	& P,St	& K,V	& P,V	& P,V	& P,St, V  &	&	& K,P, V& P,St, V & P,St	& P,St, V & P,St	&	& P,St, V &	& P	&	& St,V	&	&	& P,St	\\
(38545) & St	&	&	&	& St	&	& St	&	& St	& St	& St	& St	&	& St	&	&	&	& St	&	&	& St	\\
74873	& H	&	& H	& H	& H	& H	&	&	& H	& H	& H	& H	&	& H	&	& H	&	& H	&	&	&	\\
75654	& 	& K	&	&	& So	&	& 	& K	& So	& 	& So	& So	& So	& So	&	&	&	& 	&	&	& 	\\
81290	& 	&	&	&	& So	&	& 	&	& So	& So	& So	& So	& So	& So	&	&	&	& 	&	&	& 	\\
83041	& 	&	&	&	& So	&	& 	&	& So	& So	& So	& 	& 	& So	&	&	&	& 	&	&	& 	\\
84123	& H	& H	& H	& H	& H	&	& H	& H	& H	& H	& H	& H	& H	& H	& H	& H	& H	& H	& H	& H	& H	\\
84948A	&	&	&	& H	& H	&	& H	&	& H	& H	& H	& H	&	& H	&	& H	& H	& H	& H	&	& H	\\
84948B	&	&	&	& H	& H	&	& H	&	& H	& H	& H	& H	&	& H	&	& H	& H	& H	& H	&	& H	\\
101108	& H	&	&	&	& H	& H	& H	&	& H	& H	& H	& H	& H	& H	&	& H	&	& H	& H	&	& H	\\
105759	& 	&	&	&	& M	& 	& 	&	& 	& 	& M	& M	& M	& M	&	& 	&	& 	& 	&	& 	\\
106223	& H	&	&	&	& H	& H	& H	&	& H	& H	& H	& H	& H	& H	&	& H	& H	& H	& H	&	& H	\\
107233	& H	&	&	&	& H	& H	&	&	& H	& H	& H	& H	& H	& H,So	& H	&	&	& H	& H	& H	&	\\
109738	& 	&	&	&	& 	&	& 	&	& 	& 	& 	& 	& 	& So	&	&	&	& 	&	&	& 	\\
110411	& H	&	& H	&	& H	&	& H	&	& H,St	& H	& H	&	&	& H,St	&	&	&	& H,St	&	&	&	\\
111005	& 	&	&	&	& 	&	& 	&	& 	& 	& So	& 	& 	& So	&	&	&	& 	&	&	& 	\\
(111786) & St	&	&	& St	& St	& St	& St	&	& St	& St	& St	& St	&	& St	&	&	&	& St	&	&	& St	\\
125162	& P	& K,V	& P,V	& V	& P,V	&	& P	& K	& V	& P	& P,V	& P	&	& P,V	&	& P	&	& V	&	&	& P	\\
(141851)&	&	&	&	&	&	&	&	&	&	&	&	&	& N	&	&	&	& N	&	& N	&	\\
142703	& H	& K	&	&	& H,So	&	&	& K	& H	& H	& H,So	& H	&	& H,So	&	& H	&	& H	&	& N	& H	\\
(149303) &	&	&	&	& P	&	&	&	&	&	& P	&	&	& P	&	& P	&	&	&	&	&	\\
156954	& 	&	&	&	& 	&	& 	&	& So	& 	& So	& So	& 	& So	&	&	&	& 	&	&	& 	\\
168740	&	&	&	&	& H	&	& N	&	& H	& H	& H,So	& H	&	& H,So	&	&	&	& H	&	& N	& H	\\
170680	&	&	&	&	& H	&	&	&	&	&	& H	& H	&	& H	&	&	&	&	&	&	&	\\
171948A	&	&	& H	& H	& H	&	& H	&	&	&	&	& H	&	& H	&	&	&	&	&	&	&	\\
171948B	&	&	& H	& H	& H	&	& H	&	&	&	&	& H	&	& H	&	&	&	&	&	&	&	\\
183324	& H,St	& H,K	& H	& St	& H,St	& St	& St	& K	& H,St	& H	& H,St	& H,St	&	& H,St	&	&	&	& H,St	&	&	& St	\\
192640	& P,St	& K,V	& A,H, P,V &	& A,H, P,St, V & A & P,St& K,V& A,H, P,V & A,H& A,H, P,St, V & A,H, P,St & H	& A,H, P,St, V &	& P	&	& A,H, V & A	& A	& A,H, St \\
193256	& St	&	&	& St	& St	&	& St	&	& St	& St	& St	& St	&	& St	&	&	&	& St	&	&	& St	\\
193281	& St	& K	&	& St	& So,St	& St	& K	&	& St	&	& St	& St	&	& So,St	&	&	&	& St	&	&	& St	\\
198160	& St	&	&	& St	& St	&	& St	&	& St	&	& St	& St	&	& St	&	&	&	& St	&	&	& St	\\
198161	& St	&	&	& St	& St	&	& St	&	& St	&	& St	& St	&	& St	&	&	&	& St	&	&	& St	\\
204041	& P,St	& K	& P	&	& P,So, St& St	& P,St	& K,P	& P,St	& P,St	& P,So, St& P,So, St&	& P,So, St&	& P	&	& St	&	&	& P,St	\\
210111	& St	&	&	& St	& So,St	& St	& St	&	& So,St	& St	& So,St	& So,St	&	& So,St	&	&	&	& St	&	&	& St	\\
221756	& P	& K	& P	&	& P,St	&	& P	& K	& P	& P	& P,St	& P,St	&	& P,St	&	&	&	&	&	&	& St	\\
\hline\hline
\end{tabular}
\end{table*}

\begin{figure*}
  \resizebox{\hsize}{!}{\includegraphics{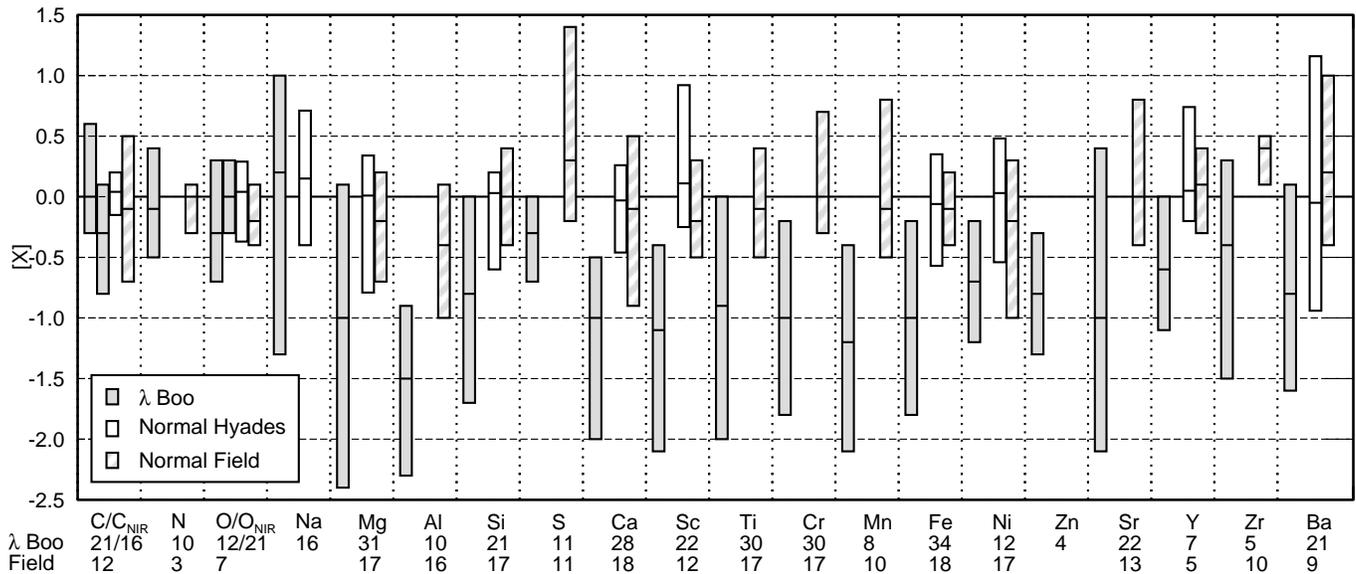}}
  \caption{Mean abundances of all analysed \LB\ stars (middle lines in grey bars), as well as highest and lowest abundances (upper and lower limit of bars). The same for normal stars in the Hyades cluster (white bars) as well as normal field stars (hashed bars). The number of available abundances for the \LB\ and field stars is given below the element name.}
  \label{mean}
\end{figure*}

Abundances for \LB\ stars have been taken from the sources listed in Table~\ref{ref_tab}.
For the construction of the abundance pattern, abundances
determined for the same element by more than one author have been
averaged. In a few cases, the abundances differ
by more than 0.6~dex and these have been dismissed from the investigation.
Also, the abundances derived by
\citet{Venn:90} for C have been discarded, because
they are based on near infrared (NIR) lines. C and O lines in this part of the
spectrum have been investigated in more detail by 
\citet{Paun:99a} and their abundances are treated as a separate set,
because they found a systematic difference between their results and that
of optical abundance analyses (see also Fig.~\ref{mean}).
Abundances are available for the following stars whose spectra should be considered
composite \citep{Fara:99}: HD\,111786 turned out to be a spectroscopic 
binary after the abundance analysis was made, and HD\,38545 and HD\,141851
are visual binaries with separations smaller than 0.2$^{\prime\prime}$ 
($\Delta m$=0.6 for HD\,38545). Indications for spectroscopic 
binarity similar to HD\,111786 were found for HD\,149303 by \citet{Paun:99a}.
The abundances derived for these stars cannot be
regarded as reliable, because the influence of the second star has not
been taken into account. Therefore they have not been included in the
investigations below and are parenthesized in Table~\ref{ref_tab}.  
The errors of the remaining abundances have been derived from the standard 
deviation in case of at least three available abundances.
From two available sources, if the two abundances were different, 
the larger of the two errors given in the references has been taken, 
and if the two abundances coincided, the smaller of the errors has been taken. 
In addition to the elements listed in Table~\ref{ref_tab}, abundances
of V, Cu, La, Nd and Eu have been derived for one star \citep{Heit:98}.

Fig.~\ref{mean} shows the abundance pattern,
which results from calculating the mean of the abundances derived for all
stars for one particular element. Also shown are the largest and smallest
occurring abundances, and the numbers of available abundances per element are
given. The mean abundance values, the standard deviations,
as well as the differences of the mean to the maximum and minimum abundances
are listed in Table~\ref{mean_tab}. When regarding only the {\em mean} abundances,
the ``\LB\ pattern'' is clearly visible, with depleted heavy elements
and nearly solar light elements, including Na and S. However, two things
are striking within this pattern: First, the difference between 
abundance determinations of C and O from optical and NIR spectra
mentioned above is obvious. Second, the abundance
of Zn, which has been found to be non-depleted in the interstellar medium
(ISM abundances will be presented in \citet{Heit:01b}), is similar to that of
the other heavy elements.
All available Zn abundances have been determined in the present paper
and in \citet{Heit:98} from the two Zn~I lines given in Table~\ref{Zn_tab}.
We regard these lines to be reliable indicators of the Zn abundance,
although Zn~I is not the dominant ionization stage in the atmospheres of 
the studied stars (the fraction of Zn~I is about 5 to 15~\%). 
NLTE effects should not play an important role because the analysed lines 
are rather weak. Furthermore, they have been used by \citet{Sned:91}
for the study of Zn abundances in disk and halo stars with metallicities
of $-$3 to $-$0.2~dex relative to the Sun and they found the abundances of
Zn to be equal to the average of other metals within $\pm$0.2~dex.
For a sample of normal late-B stars ($-0.2 \le$ [Fe/H] $\le 0.2$)
\citet{Smit:94} has determined Zn abundances from the Zn~II
resonance lines at 2025 and 2062~\AA\ and also found [Zn/Fe] = 0 $\pm$ 0.4. 

\begin{table}
\caption{Zn~I lines analysed in the four \LB\ stars HD\,84123 (1), HD\,84948A (2), HD\,84948B (3) and HD\,106223 (4).}
\label{Zn_tab}
\begin{tabular}{ccccccc}
\hline\hline
                &          &                    & \multicolumn{4}{c}{Equivalent width [m\AA]} \\
$\lambda$ [\AA] & log $gf$ & $E_{\rm low}$ [eV] & 1  & 2   & 3  & 4  \\
\hline
4722.153        & -0.338   & 4.030              & 16 &  25 & 30 & 25 \\
4810.528        & -0.137   & 4.078              & 22 &  20 &    & 40 \\
\hline\hline
\end{tabular}
\end{table}

Regarding the {\em minimum} and {\em maximum} abundances,
there is a large scatter in abundance from star to star.
More than half of the 15 elements which are {\em on the average} depleted by
$-$0.4 to $-$1.5~dex have solar abundance in some stars
and Sr and Zr can even be overabundant.
Na shows both heavy element like depletions and the largest overabundances
of all elements.

To examine the significance of these variations, a similar diagram
has been produced for 33 normal A and F dwarfs, which are members of the
Hyades cluster and whose effective temperatures and \Vsini\ values lie in the 
same range as that of the \LB\ stars, although the sample is biased towards
low temperatures (the highest temperature being 8300~K). 
Their abundances have been taken from \citet{Vare:99} 
and the mean values and abundance ranges are also plotted in Fig.~\ref{mean}.
A comparison reveals that for almost all elements the abundance ranges
are significantly larger in the \LB\ stars than in the normal stars,
in particular for Na and Mg. They are only comparable to that of the normal stars for Ni,
Y and Ba. Additionally, there are differences in the symmetries of the
abundance distributions for Na, Si, Sc and Y.

The third set shown in Fig.~\ref{mean} represents the mean abundances 
for a smaller and somewhat more inhomogeneous sample of normal 
B to F stars in the galactic field, which
is biased towards high temperatures (more than half of the stars are hotter than
10000~K). They have been extracted from a series of papers 
\citep{Adel:80,Adel:86c,Adel:87a,Adel:91a,Adel:91b,Adel:92,Adel:94a,Adel:94b,Adel:96a,Adel:97,Cali:97,Adel:98}. 
On the whole, their abundance pattern looks similar to that of the Hyades stars,
although the variations of the {\em mean} abundances around zero are larger.
For C and Ca, the abundance ranges are larger and
comparable to that of the \LB\ stars. On the other hand, the ranges for
Sc, Fe and Ba are smaller, but that may be due to the smaller sample size.
Note that the mean abundance of Al is the lowest in the normal field
stars (almost $-$0.5~dex) as well as in the \LB\ stars ($-$1.5~dex), 
and that the abundances of S in the normal field stars reach remarkably high values. 
The largest mean abundance occurs
for Zr. It is thus possible that this sample of normal stars is contaminated
with mildly chemically peculiar stars.

\begin{table}
\caption{Mean abundances of all analysed \LB\ stars (second column), as well as standard deviations (third column) and difference to highest and lowest abundances (fourth column).}
\label{mean_tab}
\begin{tabular}{lrrr}
\hline\hline\\
C  &    0.0 & $\pm$0.2 & $^{\textstyle +0.6}_{\textstyle -0.3}$ \\[2mm]
N  & $-$0.1 & $\pm$0.3 & $^{\textstyle +0.5}_{\textstyle -0.4}$ \\[2mm]
O  & $-$0.3 & $\pm$0.3 & $^{\textstyle +0.6}_{\textstyle -0.4}$ \\[2mm]
Na &    0.2 & $\pm$0.6 & $^{\textstyle +0.8}_{\textstyle -1.5}$ \\[2mm]
Mg & $-$1.0 & $\pm$0.7 & $^{\textstyle +1.1}_{\textstyle -1.4}$ \\[2mm]
Al & $-$1.5 & $\pm$0.6 & $^{\textstyle +0.6}_{\textstyle -0.8}$ \\[2mm]
Si & $-$0.8 & $\pm$0.5 & $^{\textstyle +0.8}_{\textstyle -0.9}$ \\[2mm]
S  & $-$0.3 & $\pm$0.2 & $^{\textstyle +0.3}_{\textstyle -0.4}$ \\[2mm]
Ca & $-$1.0 & $\pm$0.4 & $^{\textstyle +0.5}_{\textstyle -1.0}$ \\[2mm]
Sc & $-$1.1 & $\pm$0.4 & $^{\textstyle +0.7}_{\textstyle -1.0}$ \\[2mm]
\hline\hline
\end{tabular}
\hspace{3mm}
\begin{tabular}{lrrr}
\hline\hline\\
Ti & $-$0.9 & $\pm$0.5 & $^{\textstyle +0.8}_{\textstyle -1.1}$ \\[2mm]
Cr & $-$1.0 & $\pm$0.4 & $^{\textstyle +0.8}_{\textstyle -0.8}$ \\[2mm]
Mn & $-$1.2 & $\pm$0.5 & $^{\textstyle +0.8}_{\textstyle -0.9}$ \\[2mm]
Fe & $-$1.0 & $\pm$0.4 & $^{\textstyle +0.8}_{\textstyle -0.8}$ \\[2mm]
Ni & $-$0.7 & $\pm$0.3 & $^{\textstyle +0.5}_{\textstyle -0.5}$ \\[2mm]
Zn & $-$0.8 & $\pm$0.4 & $^{\textstyle +0.5}_{\textstyle -0.5}$ \\[2mm]
Sr & $-$1.0 & $\pm$0.6 & $^{\textstyle +1.4}_{\textstyle -1.1}$ \\[2mm]
Y  & $-$0.6 & $\pm$0.4 & $^{\textstyle +0.6}_{\textstyle -0.5}$ \\[2mm]
Zr & $-$0.4 & $\pm$0.7 & $^{\textstyle +0.7}_{\textstyle -1.1}$ \\[2mm]
Ba & $-$0.8 & $\pm$0.4 & $^{\textstyle +0.9}_{\textstyle -0.8}$ \\[2mm]
\hline\hline
\end{tabular}
\end{table}


\section{Conclusions}
We have presented new abundances of up to 15 elements for twelve \LB\ stars,
which include five Y abundances. This element has up to now been studied only for
two stars: HD\,84123 \citep{Heit:98} and 29\,Cyg \citep{Adel:99}. In these stars, the 
Y abundances are enhanced with respect to that of most other heavy elements.
The same is found for all stars presented here. The other heavy elments show underabundances
of varying degree, up to about $-$2~dex, and the light elements
C, O and Na appear moderately over- or underabundant, except for
HD\,171948A/B. In this SB2 system [C] and [O] are less than $-$0.5~dex, but
this is still high compared to the Fe abundance of $-$1.6~dex.

The results of this work consolidate the membership of all
candidate stars to the group of \LB\ stars, nearly doubling the amount
of spectroscopically investigated members of this group of peculiar stars. 

In addition, we have collected abundances for 34 \LB\ stars from the literature.
They were used to construct a ``mean \LB\ abundance pattern'', which can be
described as follows:
\begin{itemize}
\item The iron peak elements from Sc to Fe as well as Mg, Si, Ca, Sr and Ba
are depleted by about $-$1~dex relative to the solar chemical composition.
The mean abundance of Zn is similarly low, although its condensation temperature
is similar to that of S and should prevent the depletion of this element
by chemical separation within an accretion scenario.
\item Al is slightly more depleted ($-$1.5~dex) and Ni, Y and Zr are slightly
less depleted.
\item The abundances of the light elements C, N and O as well as S lie around
the solar value, which is not surprising because this is part of
the definition of the \LB\ group. 
The mean abundance of Na is also solar, but the star-to-star scatter is much 
larger for this element ($\pm$1~dex). 
\item The star-to-star scatter is twice as large as for a comparable sample
of normal stars for all elements except Ba, which means that for more than
half of the heavy elements at least one star is included in the sample,
for which the abundance of one or more of these elements is solar.
\end{itemize}

In contrast to the results of \citet{Paun:99a}, C is on average {\em more} abundant than O when
the abundances are determined from optical lines.
The enhanced abundance of Na has been already noted by \citet{Stue:93},
who derived a mean abundance of +0.6~dex for his sample of stars, and by \citet{Paun:99b}.

This description suggests the existence of a separate
chemically peculiar group of ``\LB\ stars'' with a characteristic abundance pattern.
On the other hand, the scatter of abundances for each element
indicates that the ``\LB\ group'' is rather inhomogeneous.
Furthermore, the comparison to normal stars is difficult because the sample of ``normal'' 
main sequence stars with known abundances and parameters similar to that
of the \LB\ stars is rather limited.


\begin{acknowledgements}
The author would like to thank E. Paunzen, M. Weber and R.O. Gray 
for providing observations and W.W. Weiss for his continuous
support and many useful discussions.
Thanks go to the referee, K.A. Venn, whose comments have helped to greatly
improve the paper.
This research was carried out within the working group {\em
Asteroseismology--AMS}, supported by the Fonds zur F\"orderung der 
wissenschaftlichen Forschung (project {\sl S\,7303-AST}).
Use was made of the Simbad database, operated at CDS, Strasbourg, France.
\end{acknowledgements}

\bibliographystyle{aa}
\bibliography{LB}

\end{document}